\documentclass[runningheads]{llncs}
\usepackage{graphicx}
\usepackage{amsmath}
\usepackage{appendix}
\usepackage{amssymb}
\usepackage{bbding}
\usepackage{subcaption}
\usepackage{makecell}
\newcommand{{\our}}{\textsc{Chat-Rec}}
\begin{document}
\title{Chat-REC: Towards Interactive and Explainable LLMs-Augmented Recommender System}
\titlerunning{Chat-REC: LLMs-Augmented Recommender System}

\author{
Yunfan Gao\inst{1} \and
Tao Sheng\inst{1} \and
Youlin Xiang\inst{1} \and
Yun Xiong\inst{1} \and
Haofen Wang\inst{2} \and
Jiawei Zhang\inst{3}
}

\authorrunning{Gao et al.}
%
\institute{
Shanghai Key Laboratory of Data Science, School of Computer Science, Fudan University, Shanghai, China \\
\email{yufan1602@163.com}\\
\email{tsheng16@fudan.edu.cn} \\
\email{21210240365@m.fudan.edu.cn} \\
\email{yunx@fudan.edu.cn} \and
College of Design and Innovation, Tongji University, Shanghai, China\\
\email{carter.whfcarter@gmail.com} \and
IFM Lab, Department of Computer Science, University of California, Davis, CA, USA\\
\email{jiawei@ifmlab.org}
}
\maketitle              
\begin{abstract}
Large language models (LLMs) have demonstrated their significant potential to be applied for addressing various application tasks. However, traditional recommender systems continue to face great challenges such as poor interactivity and explainability, which actually also hinder their broad deployment in real-world systems. To address these limitations, this paper proposes a novel paradigm called {\our} (ChatGPT Augmented Recommender System) that innovatively augments LLMs for building conversational recommender systems by converting user profiles and historical interactions into prompts. {\our} is demonstrated to be effective in learning user preferences and establishing connections between users and products through in-context learning, which also makes the recommendation process more interactive and explainable. What's more, within the {\our} framework, user's preferences can transfer to different products for cross-domain recommendations, and prompt-based injection of information into LLMs can also handle the cold-start scenarios with new items. In our experiments, {\our} effectively improve the results of top-k recommendations and performs better in zero-shot rating prediction task. {\our} offers a novel approach to improving recommender systems and presents new practical scenarios for the implementation of AIGC (AI generated content) in recommender system studies.

\keywords{LLMs \and Recommender System \and Prompt Engineering}
\end{abstract}
\section{Introduction}
With the scaling of model and corpus size, LLMs (Large Language Models) have shown remarkable capabilities, such as complex inference, knowledge inference, and external robustness \cite{chowdhery2022palm,fu2022gptroadmap}. These capabilities, referred to as Emergent Abilities, only become apparent after reaching a specific threshold of model parameters \cite{Wei2022EmergentAO}. The emergence of LLMs has brought about a paradigm shift in research. Previously, applying models to downstream tasks typically involved adjusting model parameters through backpropagation. However, the latest development of LLMs \cite{touvron2023llama} has enabled both researchers and practitioners to facilitate learning during the forward process by constructing prompts, namely In-Context Learning (ICL) \cite{brown2020language}.
In addition, the adoption of techniques such as Chain-of-Thought \cite{Wei2022ChainOT} and Instruct Learning \cite{Wei2021FinetunedLM} has further harnessed the reasoning capabilities and task generalization abilities of LLMs, thereby promoting their application across various domains.

In the era of big data, manual information searching has become infeasible and recommender systems have been widely deployed for automatically inferring people's preference and providing high-quality recommendation services. However, due to the great limitations and drawbacks in both model design and data distribution biases, most existing recommender systems still have great performance in their real-world deployment. One of the primary constraints is their poor interactivity, explainability, and lack of feedback mechanisms. Another limitation is the cold start problem, which makes it difficult to provide accurate recommendations for both new items and new users. Lastly, current recommender systems face challenges in making recommendations across multiple domains \cite{zhu2021cross}. In many recommendation tasks, in order to obtain the required background or general knowledge, an external library or knowledge graph needs to be set up for retrieval \cite{xu2021fusing} or multi-task learning needs to be trained on augmented data \cite{khashabi2020unifiedqa}. LLMs offer a promising solution to these challenges. They can generate more natural and explainable recommendations, solve the cold start problem, and make cross-domain recommendations. Additionally, LLMs have stronger interactivity and feedback mechanisms, which enhance the overall user experience. By leveraging internal knowledge, LLMs can improve the performance of recommender systems without relying on external retrievers \cite{yu2023generate}.

Applying LLMs for addressing the recommendation tasks has received several preliminary research experimental trials already \cite{liu2023pre,geng2022recommendation,zhang2021language}. Recommender system tasks are formulated as prompt-based natural language tasks, where user–item information and corresponding features are integrated with personalized prompt templates as model inputs. However, in the current research, LLMs are still involved in training as part of the model. 

In this paper, we introduce a novel approach to learning conversational recommender systems augmented by LLMs, which possess both interactive and explainable capabilities. We present a paradigm called {\our} (ChatGPT Augmented Recommender System) that does not require training and instead relies solely on in-context learning, resulting in more efficient and effective outcomes.
With LLM-enhanced recommender system, it is beneficial to learn users' preferences during the conversation. After each step of the conversation, the user's preferences can be further drilled down to update the candidate recommendation results. In addition, users' preferences between products are linked, allowing for better cross-domain product recommendations. We conducted recommendation and rating tests on real-world datasets and experimental results show that {\our} achieves significant improvements. {\our} sheds light on a promising technical route for the application of conversation AI such as ChatGPT in multiple recommendation scenarios.


Our contributions are summarized as follows:
\begin{itemize}
    \item We introduce a novel and effective paradigm called {\our}, which combines traditional recommender systems with LLMs through prompts, leveraging LLMs' ability to learn from context.
    \item {\our} employs LLMs as a recommender system interface, enabling multi-round recommendations, enhancing interactivity and explainability.
    \item We evaluate our method on real-world datasets for top-k recommendation and rating prediction tasks and experiments demonstrate the effectiveness of {\our}.
\end{itemize}

\section{Related Work}

\subsection{Augmented Language Models}
Augmented Language Models (ALMs) are a new research direction that aims to overcome the limitations of traditional Language Models (LMs) \cite{devlin2018bert,brown2020language,chowdhery2022palm} by equipping them with reasoning skills and the ability to use external tools, which has served millions of users, such as the coding assistant Copilot \cite{chen2021evaluating}, or more recently ChatGPT based on GPT3.5 and GPT4\footnote{https://openai.com/blog/chatgpt/}. Reasoning is defined as breaking down complex tasks into simpler subtasks that the LM can solve more easily by itself or with the help of tools \cite{lecun2022path,schick2023toolformer,parisi2022talm}, while tools are external modules that the LM can call to augment its context. ALMs can use these augmentations separately or in combination to expand their context processing ability and outperform most regular LMs on several benchmarks. ALMs can learn to reason, use tools, and even act, while still performing standard natural language tasks. This new research direction has the potential to address common limitations of traditional LMs such as interpretability, consistency, and scalability issues. By jointly discussing reasoning and tools, and tools and actions, ALMs can solve a broad range of complex tasks without heuristics, thus offering better generalization capabilities.

\subsection{NLP for Recommendation}
The field of recommender systems has had a long-standing relationship with natural language processing (NLP) techniques, especially when pre-trained language models (PLMs) comes out, which improve the performance of recommender systems and explainability \cite{chen2019personalized,li2020generate,li2021personalized}. PLMs are language models that have learned universal representations on large corpora in a self-supervised manner, and the learned representations can be beneficial to a series of downstream NLP tasks. In the recommendation domain, PLMs can help alleviate the data sparsity issue, which is a major performance bottleneck of current deep recommendation models. By extracting and transferring knowledge from pre-trained models learned by different PLM-related training paradigms, researchers aim to improve recommendation performance from various perspectives, such as generality, sparsity, efficiency, and effectiveness. In this vibrant field, there are open issues and future research directions that need to be explored, including the connection between PLM-based training paradigms and different input data types for recommender systems. Overall, adapting language modelling paradigms for recommendation is seen as a promising direction in both academia and industry.

\subsection{Cold-start Recommendation}
Cold start recommendation is a problem that arises in recommender systems when users or items have no prior interaction records with the system. This means that there is no data available for the system to make personalized recommendations. To address this issue, solutions have been proposed that either learn to model content features \cite{shi2019adaptive} or transfer representations from auxiliary domains \cite{yuan2021one,zhu2021cross}. The former approach focuses on learning about the characteristics of the items or users based on their content, such as text, images, or metadata. The latter approach involves leveraging information from other domains, such as social networks or product descriptions, to infer user preferences. Additionally, there are approaches that aim to quickly adapt to new domains instead of only providing recommendations for cold-start cases. A good generalization ability of recommendation models on startup cases is essential to ensure a better user experience and increased engagement. In our work, we use the reasoning and background knowledge of LLMs to enhance the performance of recommender systems for cold start scenarios.

\section{Method}
\vspace{-10pt}
\begin{figure}[ht]
\makebox[\textwidth][c]{\includegraphics[width=1.01\textwidth]{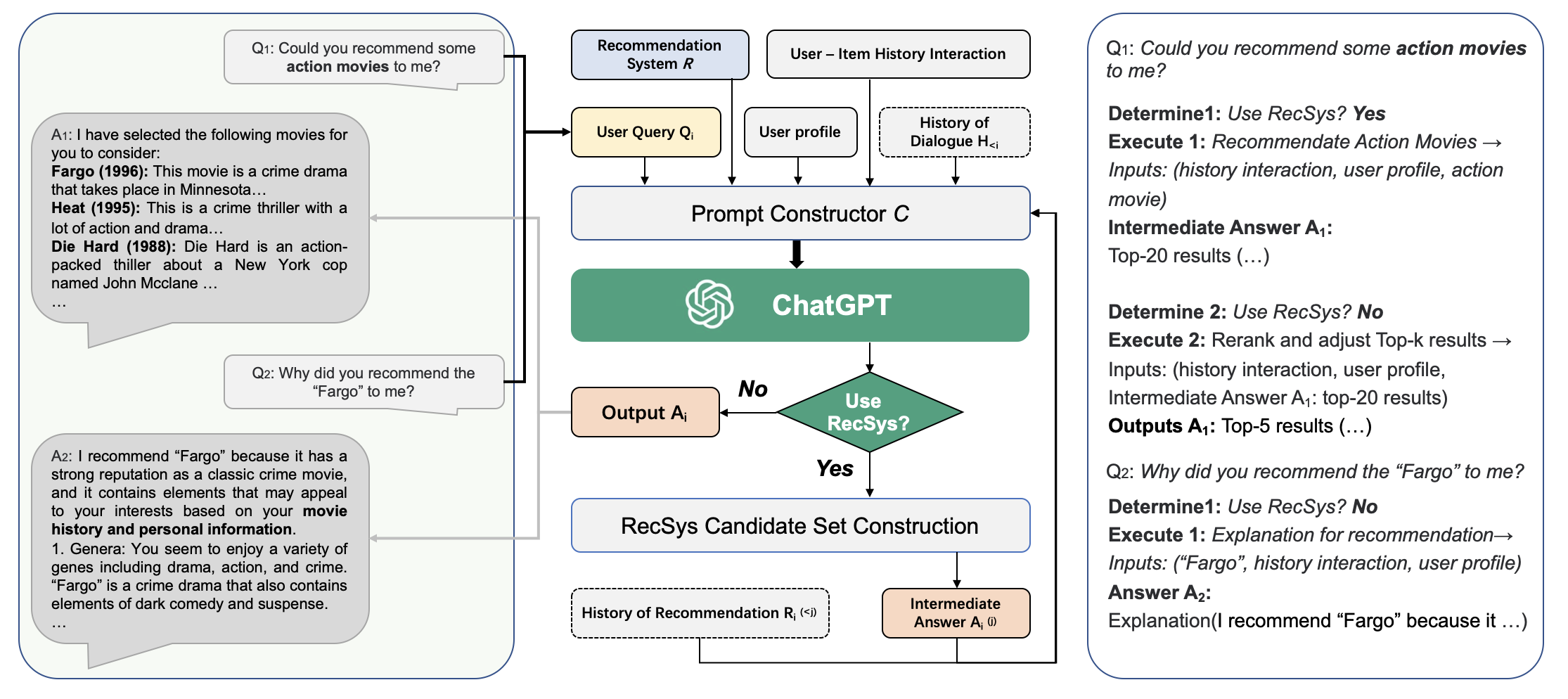}}
\caption{Overview of {\our}. The left side shows a dialogue between a user and ChatGPT. The middle side shows the flowchart to how {\our} links traditional recommender systems with conversational AI such as ChatGPT. The right side describes the specific judgment in the process.} 
\label{overview}
\end{figure}

\subsection{Bridge Recommender Systems and LLMs}

Recommender systems are designed to suggest items to users based on their preferences and behavior. Traditionally, these systems have relied on user data such as clickstream and purchase history to make recommendations. However, NLP techniques have proven to be valuable in expanding the scope of recommender systems beyond traditional user data.

NLP techniques can be used to analyze user-generated content such as reviews and social media posts to gain insights into user preferences and interests. LLMs can also be used to generate natural language responses to user queries, improving the overall user experience and engagement.

To bridge recommender systems and LLMs, we propose an enhanced recommender system module based on ChatGPT, a large language model trained by OpenAI. As the Fig. \ref{overview} shows, the module takes as input user-item history interactions, user profile, user query $Q_i$, and history of dialogue $H_{<i}$ (if available, and the notation $_{<i}$ denotes the dialogue history prior to the current query), and interfaces with any recommender system R. If the task is determined to be a recommendation task, the module uses R to generate a candidate set of items. Otherwise, it directly outputs a response to the user, such as an explanation of a generation task or a request for item details.

The prompt constructor module in the enhanced recommender system takes multiple inputs to generate a natural language paragraph that captures the user's query and recommendation information. The inputs are as follows:

\begin{itemize}
    \item User-item history interactions, which refers to the user's past interactions with items, such as items they have clicked, purchased, or rated. This information is used to understand the user's preferences and to personalize the recommendation.
    \item User profile, which contains demographic and preference information about the user. This may include age, gender, location, and interests. The user profile helps the system understand the user's characteristics and preferences.
    \item User query $Q_i$, which is the user's specific request for information or recommendation. This may include a specific item or genre they are interested in, or a more general request for recommendations in a particular category.
    \item History of dialogue $H_{<i}$, which contains the previous conversation between the user and the system. This information is used to understand the context of the user's query and to provide a more personalized and relevant response.
\end{itemize}

\begin{figure}[htbp]
\centering
\makebox[\textwidth][c]{\includegraphics[width=1.2\textwidth]{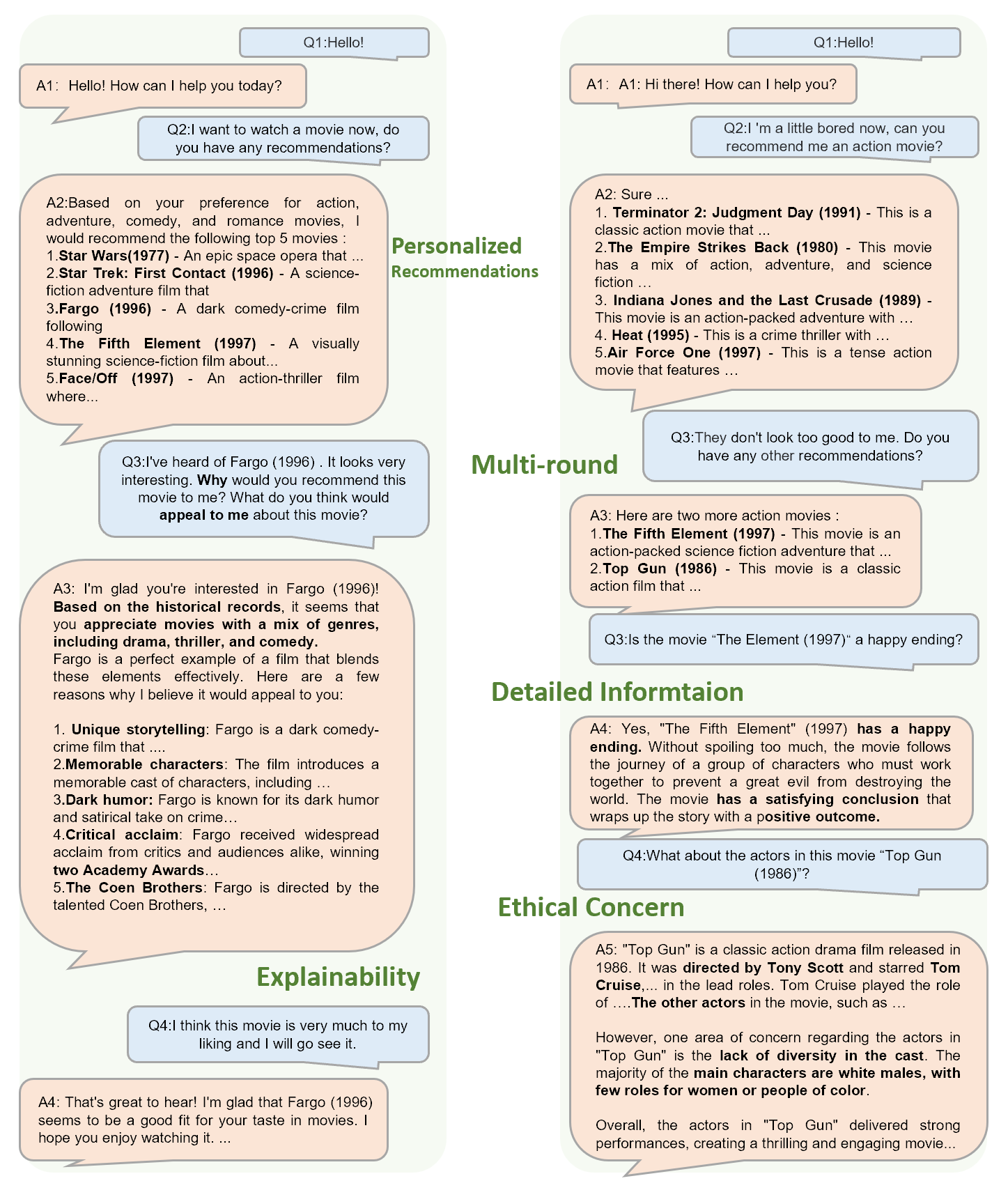}}
\caption{Case study of interactive recommendation. It shows  two conversations between different users and LLM . Where the user profile and historical users are converted into corresponding prompts for personalized recommendations, but the input of this part of the prompts is not visible to the user. The dialogue on the left shows that when a user asks why the movie was recommended, LLM can give an explanation based on the user's preferences and specific information about the recommended movie. The dialog on the right shows that {\our} can make multiple rounds of recommendations based on user feedback. Questions about the details of the movie can also be answered in a specific way. LLM also takes into account ethical and moral issues when recommending movies.} 
\label{fig:dial}
\end{figure}

As shown in Fig. \ref{fig:dial}, the {\our} framework proposed in this paper empower recommender systems with the conversational interface, which makes the interactive and explainable recommendation possible. Formally, based on the aforementioned inputs, the prompt constructor module generates a natural language paragraph that summarizes the user's query and recommendation information, and provides a more personalized and relevant response to the user's request. The intermediate answer generated by the recommender system is then used to refine the prompt constructor and generate an optimized prompt to further compress and refine the candidate set. The resulting recommendation and a brief explanation are output to the user.

For example, in the first round of Q$\&$A, the user requests action movies. The system determines that a recommendation task is needed, and executes the Recommendate Action Movies module using the input information. The intermediate answer $A_1$ contains the top-20 results, which are then reranked and adjusted in the second module using the input information to generate the final output of the top-5 results.

In the second round of Q$\&$A, the user asks why the movie ``Fargo'' was recommended. The system determines that no recommendation task is needed and instead executes the explanation for the recommendation module, using the movie title, history interaction, and user profile as inputs. The answer $A_2$ is then generated, which provides a brief explanation of the recommendation, including information about the user's general interests and the specific characteristics of the movie that may be appealing to the user.

\subsection{Recommendation Based on Candidate Set Compression}
Traditional recommender systems typically generate a small number of sorted candidate products, each with a score that reflects the system's recommendation confidence or result quality. However, considering the huge size of the product set, the performance obtained by most existing recommender systems are all way far from satisfactory, which still have a very large room for improvement.

This article proposes a method of using LLMs to improve the performance of recommender systems by narrowing down the candidate set. The recommender system generates a large set of candidate items, which can be overwhelming for the user. LLMs play several different critical roles in narrowing down the product candidate set within the system. Firstly, we convert users' profiles and historical interactions into prompts, including the item description and user rating. Secondly, LLMs are asked to summarize user preferences for items in a domain based on the above information. LLMs can learn from context and effectively capture users' background information and preferences. With this information, they can establish the relationship between product attributes and user preferences, enabling them to make better product recommendations. By utilizing in-context learning, LLMs can enhance their recommendation reasoning ability, resulting in more accurate and personalized product recommendations.

Once the LLMs have learned the user's preferences, the candidate set generated by the recommender system is provided to the LLMs. The LLMs can further filter and sort the candidate set based on the user's preferences. This approach ensures that the user is presented with a smaller, more relevant set of items, increasing the likelihood that they will find something they like.

\subsection{Cold-start Recommendations}

\begin{figure}[htbp]
\includegraphics[width=\textwidth]{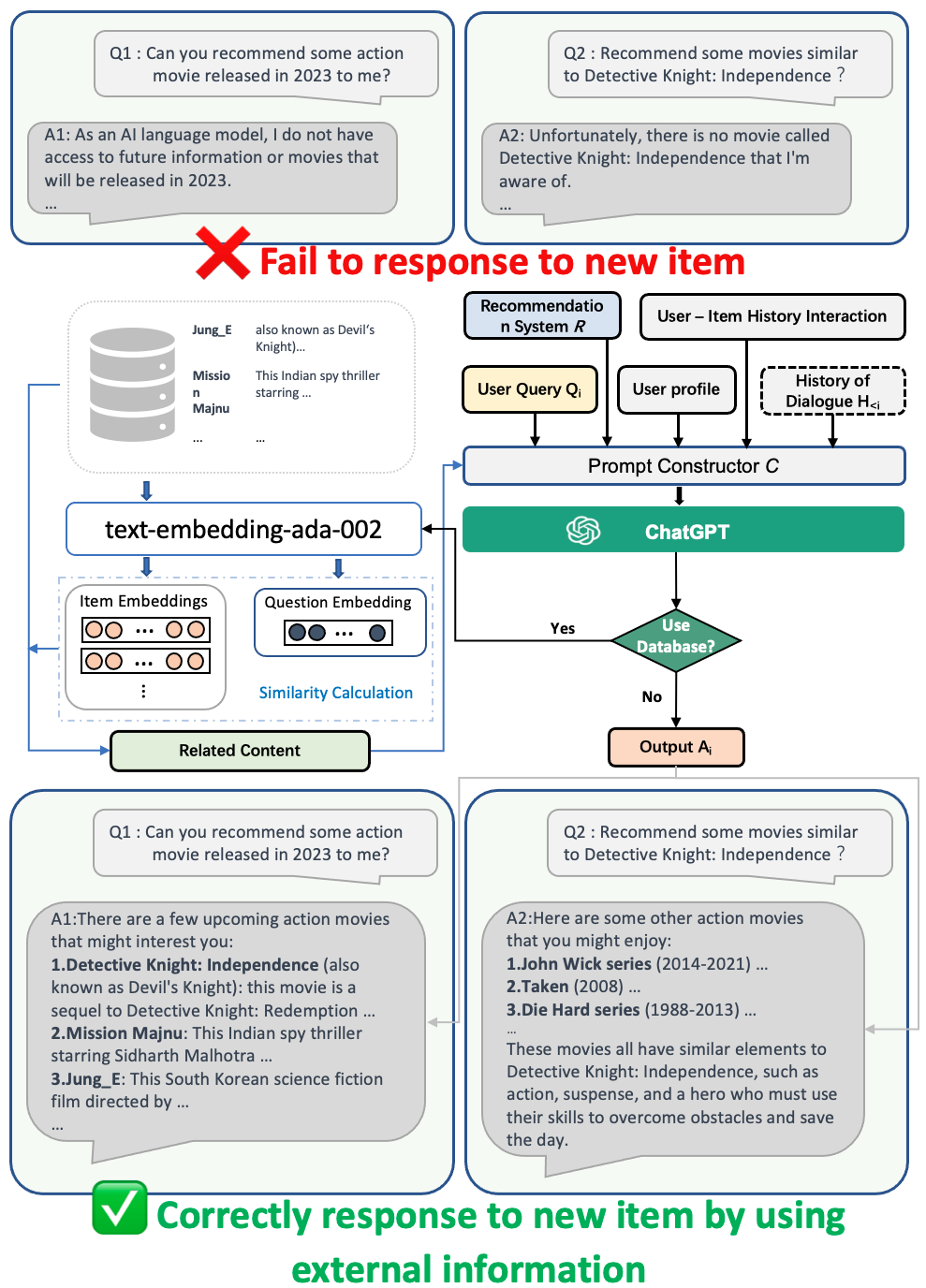}
\caption{Case Study of New Item Recommendation. The top shows that ChatGPT is unable to recommend new items beyond the timeframe of its training data. The middle part demonstrates the process of how to utilize external information about new items to enable ChatGPT to handle recommendations for new items. The bottom shows that ChatGPT can effectively handle recommendations for new items after incorporating external information. } 
\label{fig:newItem}
\end{figure}

With the textual description and profile information about the products, regardless the new products or the old ones, LLMs can effectively relate such products with each other, which provides us with the opportunity for solving the persistent cold-start recommendation problem once and for all.

For example, if a user asks for recommendations for a new movie that was released in 2021, the recommender system could use text data about the movie to generate an embedding and then calculate similarities to other movies in the system to make recommendations. This capability allows recommender systems to make relevant and accurate recommendations for new items, improving the overall user experience.

Large language models can use the vast amount of knowledge they contain to help recommender systems alleviate the cold-start problem of new items, i.e., recommending items that lack a large number of user interactions.
However, since the knowledge held by ChatGPT is limited to September 2021, ChatGPT does not cope well when encountering unknown items, such as a user requesting to recommend some new movies released in 2023 or  content related to a movie that ChatGPT is not aware of, as shown in the top part of Fig. \ref{fig:newItem}.
To address this issue, we introduce external information about new items, utilizing large language models to generate corresponding embedding representations and cache them. When encountering new item recommendations, we calculate the similarity between item embeddings and embeddings of user requests and preferences, then retrieve the most relevant item information based on the similarity and construct a prompt to input to ChatGPT for recommendation, as illustrated in the lower half of Fig. \ref{fig:newItem}. This approach allows the recommender system to work in conjunction with ChatGPT to better recommend new items, thus enhancing the user experience.

\subsection{Cross-Domain Recommendations}
\begin{figure}[ht]
\centering
\makebox[\textwidth][c]{
\includegraphics[width=\textwidth]{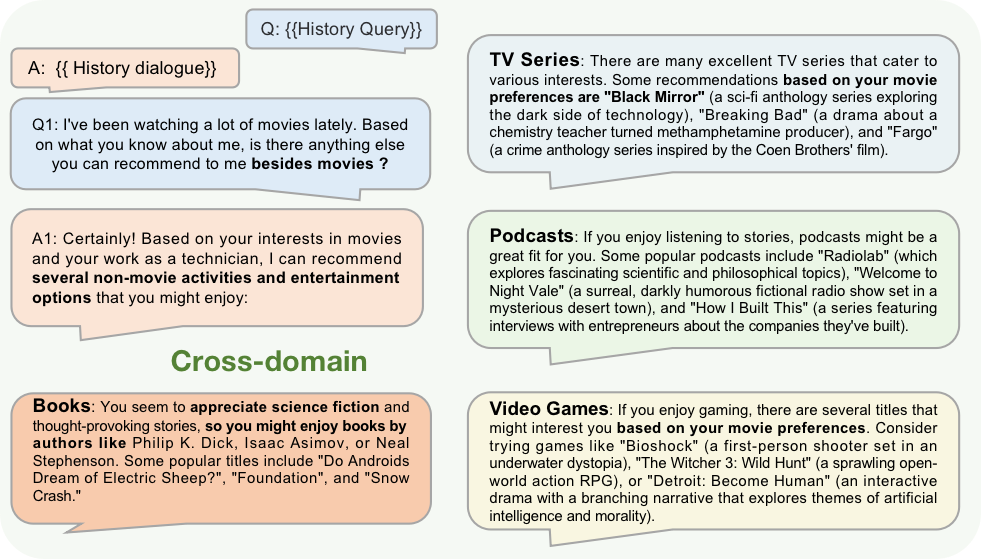}
}
\caption{Case study of cross-domain recommendation. After the conversation about the movie's recommendation is completed. The user asks LLM to recommend works other than movies. It can be seen that LLM recommends different types of works, including books, TV series Podcasts and video games, according to the user's movie preferences. This shows that LLM can migrate the user's movie preferences to items and thus achieve cross-domain recommendations.} 
\label{cross_domain}
\end{figure}
The LLMs-augmented recommender system introduced above can be used to address several challenging tasks, that are hard or even impossible to be addressed with conventional recommender systems, such as \textit{cross-domain recommendation} \cite{zhu2021cross} and \textit{cold-start recommendation} \cite{sun2020lara}. In this part, we will first talk about how to use the LLMs-augmented recommender system for the cross-domain recommendation. 

LLMs pre-trained with information across the Internet actually can serve as the multi-perspective knowledge base \cite{petroni2019language}. Besides the target product in one domain, such as movies, the LLMs not only has a broad knowledge about products many other domains, like music and books, but also understands the relations among the products across the domains mentioned above.

For example, as illustrated in Fig. \ref{cross_domain}, once the conversation regarding movie recommendations is finished, the user inquires LLM for suggestions on other types of works. LLM then proceeds to recommend a variety of options, such as books, TV series, podcasts, and video games, based on the user's movie preferences. This demonstrates LLM's ability to transfer the user's preferences from movies to other items, resulting in cross-domain recommendations. This cross-domain recommendation capability has the potential to significantly expand the scope and relevance of recommender systems.

\section{Experiment}
\subsection{Dataset and Experimental Settings}
The dataset used in our experiment is MovieLens 100K, which is a benchmark dataset of a real-world recommender system. It comprises 100,000 movie ratings provided by 943 users on a scale of 1 to 5 across 1,682 movies. Additionally, the dataset contains demographic information about the users, such as age, gender, occupation, and zip code, as well as movie information, such as title, release year, and genres. To create our experimental dataset, we randomly selected 200 users. Table \ref{dataset} provides detailed statistical information about the dataset used in the experiment.

\begin{table}[!ht]
\centering 
\caption{Details of the dataset used for evaluation.}
\setlength{\tabcolsep}{2mm}{
\begin{tabular}{cccccc}
\hline
Dataset & Users & Items & Ratings & Rating Scale &  Density  \\ 
\hline  
MovieLens 100K & 943  & 1,682 &  100,000  & [1-5] & 6.304\% \\
\hline
\end{tabular}}
\label{dataset}
\end{table}


When evaluating the performance of top-k recommendations, Precision, Recall, and Normalized Discounted Cumulative Gain (NDCG) are used. For rating prediction task, the Root Mean Squared Error (RMSE) and Mean Absolute Error (MAE) are employed as evaluation metrics. 

\subsection{Baselines}
The baseline methods studied in the experiment include both classic recommender system models and the LLMs-augmented recommender systems proposed in this paper. Detailed information about the comparison methods studied in our experiments are provided as follows:

\begin{itemize}
    \item{\textbf{LightFM}}
    is a recommendation algorithm that combines collaborative filtering and content-based methods to recommend items to users.
    \item{\textbf{LightGCN}}
    is a graph-based collaborative filtering algorithm that uses a simplified graph convolutional network (GCN) to model the user-item interactions in a recommender system.
    \item{\textbf{Item-KNN}}
    is a neighborhood-based collaborative filtering algorithm that uses the similarity between items to make recommendations to users.
    \item{\textbf{Matrix Factorization (MF)}}
    is a widely used collaborative filtering algorithm that represents users and items as latent factors in a low-dimensional space.
\end{itemize}

We select three representative models from the GPT-3 and GPT-3.5 series as LLMs in {\our}:
\begin{itemize}
    \item{\textbf{gpt-3.5-turbo}} is the most capable GPT-3.5 model and optimized for chat.
    \item{\textbf{text-davinci-003}} can do any language task with better quality, longer output, and consistent instruction-following.
    \item{\textbf{text-davinci-002}} is similar to text-davinci-003 but is trained with supervised fine-tuning instead of reinforcement learning.
\end{itemize}

The model notations, like {\our} (gpt-3.5-turbo), denote the {\our} framework built by adopting ``gpt-3.5-turbo'' as the backbone model.

\subsection{Result and Analysis}
\subsubsection{Top-5 Recommendation.}
As presented in Table \ref{tab:top5_rec}, our proposed {\our} framework has demonstrated effective improvement of traditional recommender systems in the top-k recommendation task. The NDCG scores of all three GPT-3.5 models surpassed that of LightGCN, with text-davinci-003 delivering the best result and demonstrating strong contextual learning abilities. Specifically, the precision score of 0.3240 is 6.93\% higher than that of LightGCN, while  NDCG score of 0.3802 is 11.01\% higher. However, the recall rate of 0.1404 is slightly lower than that of LightGCN by 3.51\%. It is noteworthy that the performance of gpt-3.5-turbo was slightly weaker than that of text-davinci-002. 

\begin{table}[!ht]
\centering
\caption{Results of top-5 recommendation.}
\label{tab:top5_rec}
\setlength{\tabcolsep}{1mm}{
\begin{tabular}{|c|c|c|c|}
\hline
Models  & Precision &  Recall  & NDCG \\
\hline
LightFM & 0.2830 & 0.1410 & 0.2846 \\
LightGCN & 0.3030 & \textbf{0.1455} & 0.3425 \\
{\our} (gpt-3.5-turbo) & 0.3103 &0.1279  &0.3696  \\
{\our} (text-davinci-003) & \textbf{0.3240} (+6.93\%) & 0.1404 (-3.51\%) &  \textbf{0.3802} (+11.01\%)\\
{\our} (text-davinci-002)& 0.3031  & 0.1240 & 0.3629 \\
\hline
\end{tabular}}
\end{table}
\vspace{-10pt}

\subsubsection{Rating Prediction}
As illustrated in the Table\ref{tab:rating}, {\our} outperforms traditional recommender systems in predicting movie ratings. The experimental results demonstrate that LLMs can effectively learn user preferences from user portraits and historical interactions through in-context learning, without any explicit training, and accurately predict user ratings for candidate movies. Since LightGCN is not well-suited for rating prediction tasks, it was excluded from our experimental range. Among the three GPT-3.5 models tested, text-davinci-003 achieved the best result, with an RMSE of 0.785, which is 15.86\% higher than that of Item-KNN, and an MAE of 0.593, which is 19.21\% higher. Text-davinci-002 came in second place. However, the performance of gpt-3.5-turbo was slightly weaker than that of Item-KNN. The experimental results reveal that even without relying on recommender systems, LLMs can achieve better results in predicting user preferences for specific movies. The weaker performance of gpt-3.5-turbo is due to the model's emphasis on the ability of human-computer dialogue and its trade-off of the in-context learning abilitys, which is consistent with other research conclusions. Additionally, it also can be concluded that the performance of gpt-3.5-turbo in numerical prediction tasks is weaker than that of text-davinci-003 and text-davinci-002.
\vspace{-20pt}

\begin{table}[!ht]
\centering
\setlength{\tabcolsep}{5mm}{
\caption{Results of movie rating prediction.}
\label{tab:rating}
\begin{tabular}{|c|c|c|}
\hline
 Models & RMSE & MAE \\
\hline
MF &  0.988 & 0.771 \\
Item-KNN & 0.933 & 0.734 \\
{\our} (gpt-3.5-turbo) &0.969 & 0.756 \\
{\our} (text-davinci-003) & \textbf{0.785}& \textbf{0.593} \\
{\our} (text-davinci-002) & 0.8309 &0.6215  \\
\hline
\end{tabular}}
\end{table}
\vspace{-20pt}

During  experiment, we discovered that {\our}'s most important ability is to optimize the refined candidate set of the recommender system, meaning to resort the movies that the user may like but were placed further down in the recommender system's candidate set. This requires the application of LLMs' knowledge of movies, understanding of user preferences, and the ability to reason about the matching relationship between the two. To confirm this finding, we conducted separate empirical studies and asked LLMs again, in the same conversation, about movies that appeared in the recommender system's top 5 but did not appear in LLMs' top 5. LLMs' feedback revealed that it is unlikely that the user would like the movie or it is difficult to determine whether the user would like it, with clear reasons given. The inconsistent shows that {\our}'s recommendations are entirely based on an understanding of user preferences and movie information.

\subsection{Ablation Study}

\begin{figure}[htbp]
\centering
\makebox[\textwidth][c]{
\begin{subfigure}{0.4\linewidth}
  \centering
  \includegraphics[width=\linewidth]{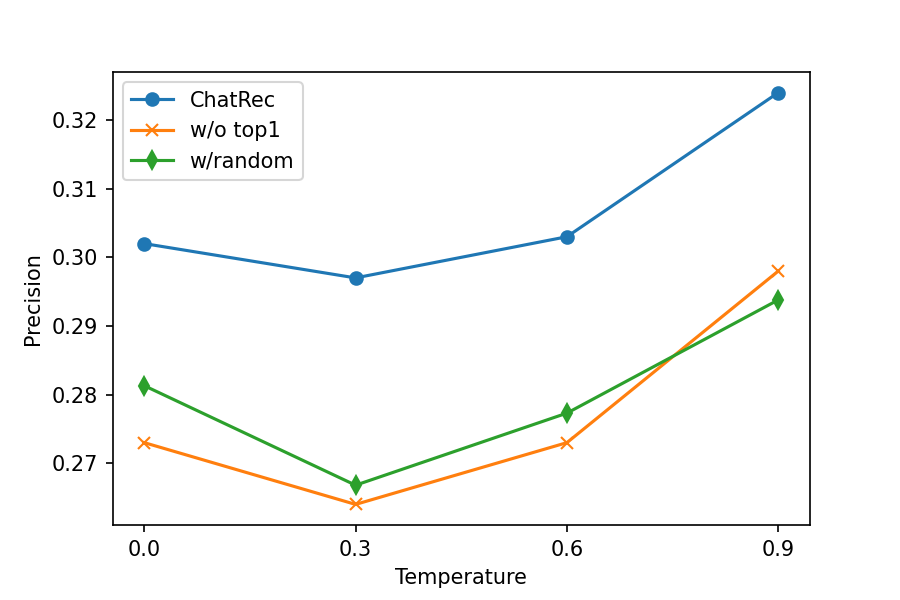}
\end{subfigure}%
\begin{subfigure}{0.4\linewidth}
  \centering
  \includegraphics[width=\linewidth]{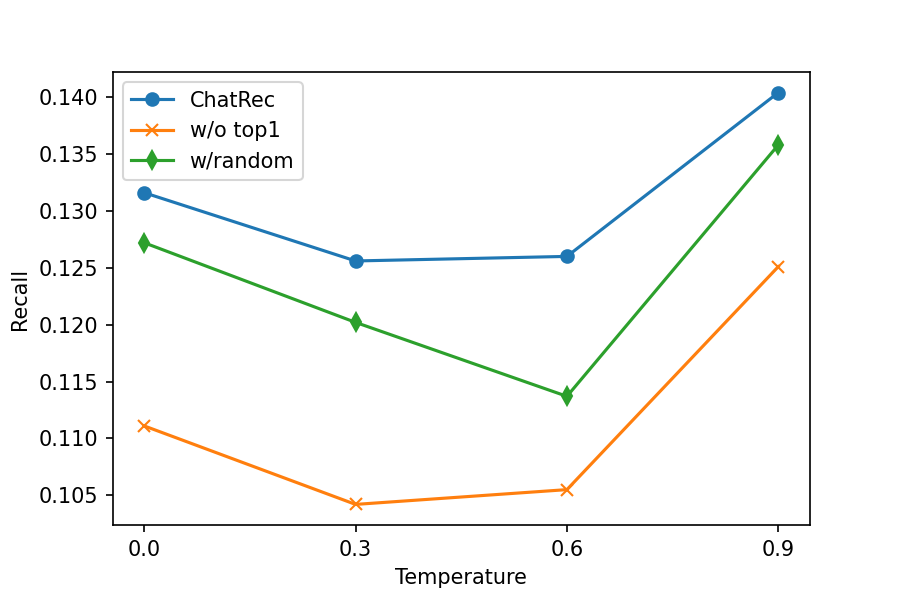}
\end{subfigure}
\begin{subfigure}{0.4\linewidth}
  \centering
  \includegraphics[width=\linewidth]{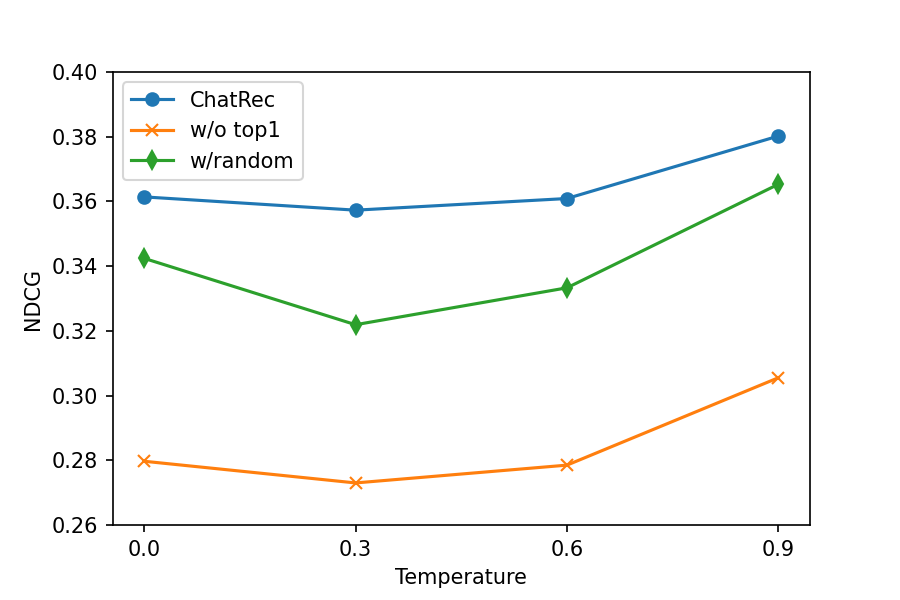}
\end{subfigure}
}
\caption{Performance on different prompt and temperature.}
\label{fig:prompt_temp}
\end{figure}

In this study, we select the text-davinci-003 model, which achieved the best results in both top-k recommendation and rating prediction, to investigate the impact of different prompts and temperatures on the model's performance. The result is shown in Fig. \ref{fig:prompt_temp}.

In the context of this study, ``w/random" refers to the random shuffling of the 20 candidate sets generated by the recommender system before being provided to LLM as the candidate set prompt input, while ``w/top1" indicates that the top 1 recommendation is not given as the initial background knowledge when constructing the prompt, but instead directly asks LLM to select 5 movies from the candidate set. The temperature parameter affects the answer generated by LLM, with lower temperatures indicating more certain answers, and higher for more random answers. All experiments, except for the experiment with a temperature of 0, used the average of 5 tests.



The results demonstrate that the effect slightly decreased after the order of the candidate set was shuffled. For example, when the temperature is 0.9, the NDCG of text-davinci-003 decreased from 0.3802 to 0.3653, representing a decrease of 3.92\%. The effect of {\our} decreased significantly when the recommender system's top 1 was missing in the prompt. For instance, when the temperature is 0.9, the NDCG of text-davinci-003 decreased from 0.3802 to 0.3055, which is a decrease of 19.65\%. This trend was observed at different temperatures, and the experiment showed that the best results could be achieved when the temperature was 0.9.

It is worth noting that the existence of the recommender system was not explicitly mentioned in {\our}'s prompt, and the function of the recommender system was merely to provide a candidate set. However, the design of the candidate set can significantly impact {\our}'s performance. Our experiment revealed that {\our}'s prompt design can effectively inject the recommender system's knowledge implicitly into LLMs. This implicit knowledge is reflected in the ranking of movies in the candidate set, and the use of Top1 as the background can further strengthen this information. This implicit knowledge can be captured by LLMs in in-context learning and can enhance the recommendation performance.



\section{Conclusion}
In this paper, we present {\our} which bridges recommender system and LLMs by converting user information and user-item interactions to prompt. We evaluated our approach in the task of top-k recommendation and zero-shot movie rating prediction. In conclusion, LLMs offer significant potential for enhancing recommender systems by improving interactivity explainability and cross-domain recommendation. In addition, prompt plays an important role, and experiments prove that implicitly expressing the knowledge in the recommender system in prompt can effectively improve the recommendation effect.


%
%
%
\bibliographystyle{splncs04}
\bibliography{main}

\appendix
\section{Implementation Details}
\subsection{Prompts}
Below, we list the prompts used in top-k recommendation and zero-shot movie rating tasks.
\begin{figure}[htbp]
\includegraphics[width=\textwidth]{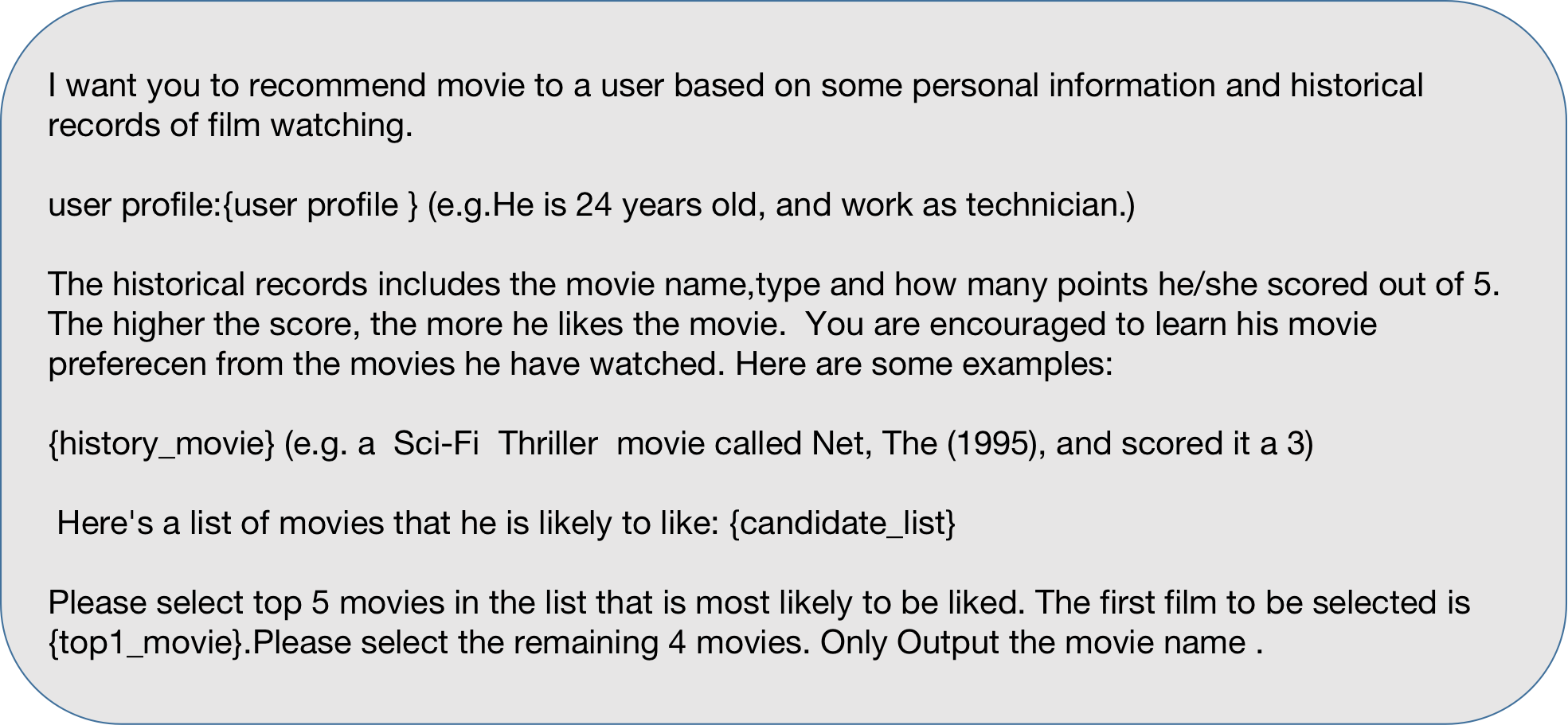}
\caption{Prompt for top-k recommendation task. } 
\label{fig:prompt1}
\end{figure}
\vspace{-20pt}
\begin{figure}[htbp]
\includegraphics[width=\textwidth]{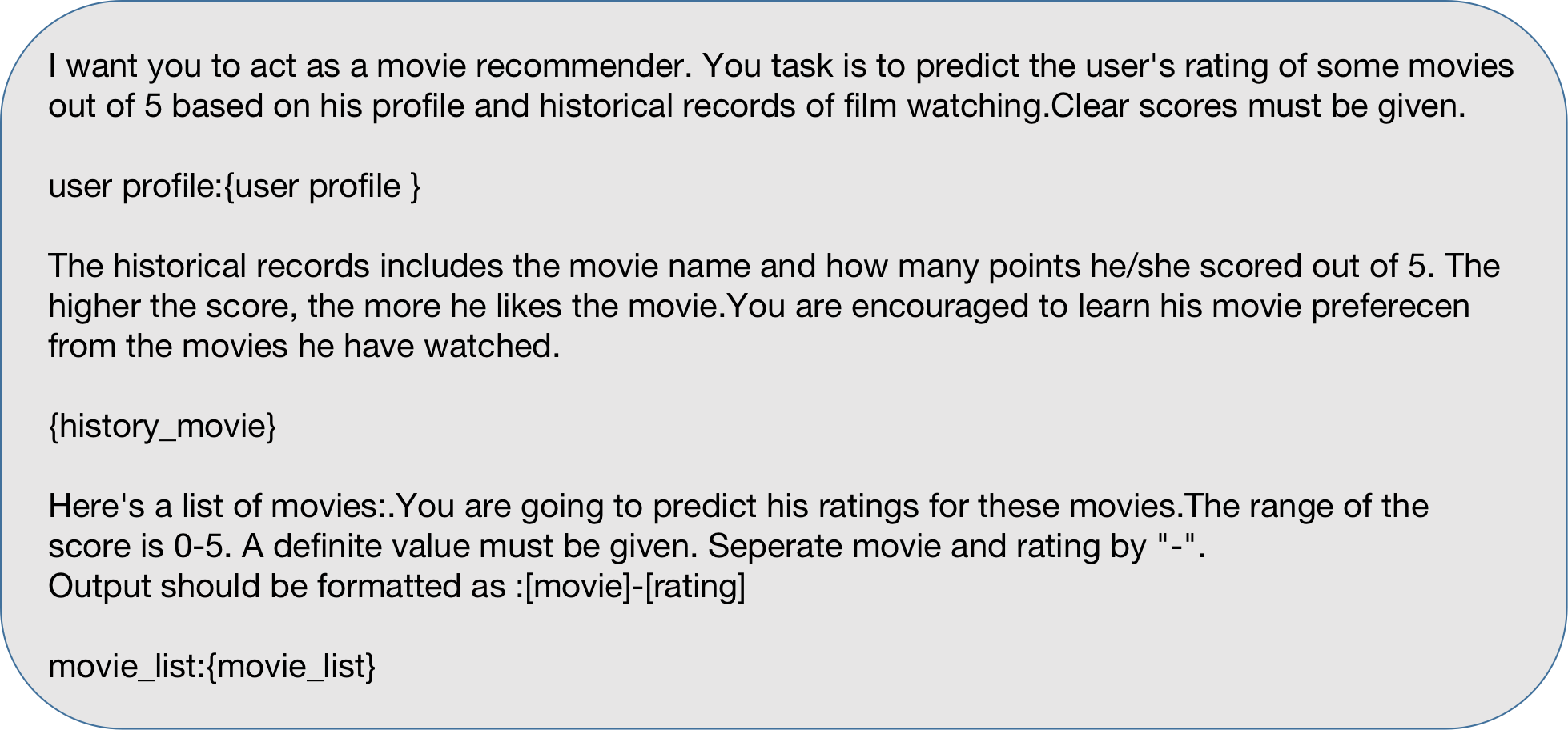}
\caption{Prompt for moving rating task. } 
\label{fig:prompt2}
\end{figure}

\subsection{Example Answers }
In fact, the LLMs do not always output answers in the format we expect every time, especially at higher temperatures. In table \ref{failed_case}, we give some failed cases while invoking LLMs' API to generate answers. During the experiment, output that does not match the format is automatically retried.

\begin{table}[htbp]
\centering 
\caption{Some cases and explanations that failed to generate canonical answers}

\setlength{\tabcolsep}{2mm}{
\begin{tabular}{p{6cm} p{4cm} c}
\hline
Example & Explanation & Correct \\
\hline
...

The current list is:
1.Toy Story (1995)
2.Fargo (1996)
3.Die Hard (1988)
4.Fish Called Wanda, A (1988)
5. Wrong Trousers, The (1993)  & The output conforms to the formatting requirements & \checkmark \\
\hline
...

The current list is:
1.The Shawshank Redemption (1994)
(It should be ``Shawshank Redemption, The(1994)")
2.A Fish Called Wanda (1988)
(It should be ``Fish Called Wanda, A (1988)")
...

& Failure to output film names in accordance with film industry norms. such as ``A" and ``The'' are not in the right place.
&\XSolidBrush  \\
\hline
...

The current list is:
1.Toy Story (1995)
2.Groundhog Day (1993)
3.Star Trek: The Wrath of Khan (1982)
4.Fargo (1996)

& Sometimes it can't output a sufficient number of movies. In this case, it only output 4 movies while sometimes may output 19 movies. & \XSolidBrush \\
\hline
...
The current list is:
a:Star Wars (1977)
a:Raiders of the Lost Ark (1981)
n:Back to the Future (1985)
m:Fargo (1996)
...
& Sometimes the id information is lost when LLM is asked to output movies in the following format [id]:[name]. & \XSolidBrush   \\

\end{tabular}}
\label{failed_case}
\end{table}

\end{document}